\documentstyle[12pt]{article}
\begin{document}

\begin{titlepage}
\begin{flushright}
UMN--TH--1122/92 \\
December 1992
\end{flushright}
\vspace{0.4in}
\begin{center}
{\Large \bf Is $\Upsilon (3S)$ a pure $S-$wave ?  \\}
\vspace{0.8in}
{\bf Sumantra Chakravarty, ~~Sun Myong Kim} \\
\vspace{0.2in}
and \\
\vspace{0.2in}
{\bf Pyungwon Ko \footnote{(pyungwon@umnhep)}} \\
\vspace{0.4in}
{\sl School of Physics and Astronomy \\
University of Minnesota \\
Minneapolis, MN 55455 \\}
\vspace{0.4in}
{\bf   Abstract  \\ }
\end{center}

Assuming the QCD multipole expansion is applicable to
hadronic transitions of $\Upsilon (3S)$ into lower level bottomonia,
we consider the possibility that $\Upsilon (3S)$ has a $D-$wave component.
This assumption leads to a natural explanation
of the $\pi \pi$ spectrum in $\Upsilon (3S) \rightarrow
\Upsilon (1S) \ \pi \pi$.
Consequences of this assumption on other hadronic and radiative
transitions of $\Upsilon (3S)$ are also discussed in the  same context.

\vspace{.3in}

\end{titlepage}

\baselineskip=30pt
\noindent
{\Large \bf  1. Introduction}

\vspace{.2in}

It has been suggested recently by two of us ( S.C. and P.K. )
that the $\pi \pi$ spectrum in
$\Upsilon (3S) \rightarrow \Upsilon (1S) \pi   \pi$ can be explained
by including a $D-$wave amplitude for the dipion
system \cite{chakrako}.
The most general amplitude for a spin--1 particle decaying into
another spin--1  particle with the emission of two pions is given by
\begin{equation}
{\cal M} = A_{0} ~ \epsilon^{\mu} \epsilon^{' \nu } ~\left[
( q^{2} + B\ E_{1} E_{2} + C\ m_{\pi}^{2} ) \ g_{\mu \nu}
+ D \ ( p_{\mu} p_{\nu}^{'} + p_{\nu} p_{\mu}^{'} ) \ \right],
\label{ampone}
\end{equation}
in the lowest order in pion momenta expansion.
Here, $\epsilon$ and $\epsilon^{'}$ are the polarization vectors of the
initial
and the final $\Upsilon$'s, $p$ and $p^{'}$ are the 4-momenta of two pions,
$q^{2} = ( p + p^{'} )^{2} \equiv m_{\pi \pi}^2 = s_{\pi}$, and $E_1$ and
$E_2$ are the energies of each pion in the rest frame of the initial
$\Upsilon$.
Two sets of parameters give the best fit to the $m_{\pi \pi}$ distribution
in $\Upsilon (3S) \rightarrow \Upsilon (1S) \pi \pi$ \cite{chakrako}.
Various angular distributions of the decay products in $e^{+} e^{-}
\rightarrow
\Upsilon (3S) \rightarrow \Upsilon (1S) \pi \pi$
are predicted in Ref. \cite{chakrako},
and these      need to be verified by future experiments.

However, the reason for $D \neq 0$ in (\ref{ampone}) was not
clearly discussed
in Ref. \cite{chakrako}.  Two possibilities were briefly mentioned :
either  a $D-$wave admixture in $\Upsilon (3S)$ or, a  breakdown of
QCD multipole expansion for hadronic transitions of $\Upsilon (3S)$.
It is our purpose to explore the first possibility in detail.
Since QCD multipole expansion enables us to understand hadronic
transitions between heavy quarkonia other than $\Upsilon (3S)$,
it is desirable to try to understand the amplitude (\ref{ampone})
in the same framework.
If this is possible,  then other hadronic transitions of
$\Upsilon (3S)$ can be  studied in the same context.
We note that theoretical predictions on hadronic
transitions of $\Upsilon (3S)$  in the literature are not reliable,
since they do not correctly describe  $\Upsilon (3S) \rightarrow
\Upsilon (1S) \pi
\pi$.
If our predictions are in serious contradiction with the experiments,
then we may have to conclude that
the QCD multipole expansion breaks down in case of $\Upsilon (3S)$.

This work is organized as follows.
In Section 2, the amplitude (\ref{ampone})  is interpreted
in the framework of QCD multipole expansion.
It is found that results in Ref.~\cite{chakrako}
can be readily obtained,
once $\Upsilon (3S)$ is assumed to be a mixture of $S-$ and $D-$waves with
a mixing angle $\phi$ :
\begin{equation}
| \Upsilon (3S) \rangle = \cos \phi ~|3S \rangle + ~\sin \phi ~| D \rangle.
\label{mixing}
\end{equation}
Consequences of this assumption on other decays of $\Upsilon (3S)$ are
then explored in detail.
First of all,   it turns out that the current upper limit on
$B (\Upsilon (3S) \rightarrow \Upsilon (1S) + \eta)$ selects P2 from
the two sets of parameters
of Ref.~\cite{chakrako}.
In Section 3, various radiative transitions of $\Upsilon (3S)$
are considered.
There, a tight constraint on the $D-$wave mixing arises from
electric dipole
radiative transitions $\Upsilon (3S) \rightarrow \chi_{bJ} (2P) +
\gamma$.
In the presence of a $D-$wave component in $\Upsilon (3S)$,
some new and interesting radiative decays appear.
It can affect the decay rate of
$\Upsilon (3S) \rightarrow \eta_{b} + \gamma$,
and allows  the following  cascade transitions :
\begin{equation}
\Upsilon (3S) \stackrel{\gamma}{\longrightarrow}~^{1}D_{2}
\stackrel{\gamma}{\longrightarrow}~
h_{b} (1P) \stackrel{\gamma}{\longrightarrow}~\eta_{b}.
\label{cascade}
\end{equation}
Besides these decays, $\Upsilon (3S) \rightarrow h_{b} (1P) + \pi^{0}$
and  $\Upsilon (3S) \rightarrow h_{b} (1P) \
\pi \pi$ are also interesting,  and the $D-$wave contributions to these
processes  are considered in Section 4.
For these decays,  we adopt the approach proposed by Voloshin \cite{v1},
which correctly  predicts the ratio of the charmonium
$^{1}P_1$ state decaying into $J/\psi + \pi^0$ and $J/\psi + \pi \pi$
\cite{e760}.
All of these decays  reach a spin--singlet $P-$wave state, $h_{b} (1P)$,
that is hard to produce  in the $e^+ e^-$ annihilation.
$h_{b} (1P)$ can be a source of a spin--singlet
$S-$wave state ($\eta_b$) through electric dipole radiative transition,
$h_{b} (1P) \rightarrow \eta_{b} + \gamma$.
Finally, our results are summarized in Section 5.

In the following,
the absolute decay rate or its lower/upper bound is derived for
each decay process.  It depends on the mixing angle $\phi$ and
quarkonium matrix elements of operators, $r$ and $r^2$, where $\vec{r}$
is the spatial separation of $b$ and $\bar{b}$.
The matrix element of $r^2$
between $\Upsilon (3S)$ and $\Upsilon (1S)$ can be  directly extracted from
the spectrum and the absolute decay rate of $\Upsilon (3S)
\rightarrow \Upsilon (1S) \ \pi \pi$, and gives
information that is independent of specific potential models.
Absolute decay rates of $\Upsilon (3S) \rightarrow
\chi_{bJ} (2P) + \gamma$ give useful
information on the matrix element of $r$ between $\Upsilon (3S)$ and
$\chi_{b}(2P)$ and the mixing angle $\phi$.
Other unknown quarkonium matrix elements will be fixed by the results
from potential model calculations.
We use $m_{b} = 4.8$ GeV in this work.  This induces some uncertainty
less than $\sim 20 \%$  in the numerical estimates of $1/ m_{b}^{2}$.
Finally, some of our results in Sections 3 and 4  show explicit
dependence  on $G_{8}$,
the  Green's function of  the color octet $b\bar{b}$ states
(defined in (\ref{greenfunction}).)
These results should be regarded  as being order--of--magnitude
estimates  because of the approximation  we will make about
$G_8$.

\vspace{.4in}
\noindent
{\Large \bf 2. Hadronic transitions of $\Upsilon (3S)$ into $\Upsilon
(1S)$}

\vspace{.2in}

Let us begin with the $\pi \pi$ spectrum in $\Upsilon (3S)
\rightarrow \Upsilon (1S) \pi \pi$.
In QCD multipole expansion,
this process occurs through $E1-E1$ multipole interaction,
where the $E1$ interaction Hamiltonian of quarkonium with a gluon is
\cite{vz}
\begin{equation}
H_{int} (E1) = - {1\over 2}~g \xi^{a} \ r_{i} E_{i}^{a} (0).
\label{e1}
\end{equation}
Here, $g \equiv (4 \pi \alpha_{s})^{1/2}$ is the $SU(3)_{c}$ gauge
coupling constant, $\vec{r}$ is the relative position of the quark ($b$)
and the antiquark ($\bar{b}$), and $\xi^{a} \equiv t_{b}^{a} -
t_{\bar{b}}^{a}$ is the difference of the $SU(3)_{c}$ color generators that
act on the quark and the antiquark, respectively.  When acting
bewteen colorless  states, $\xi^a$ satisfies
\begin{equation}
\langle \ {\rm color singlet} \ | \ \xi^{a} ~\xi^{b} \ | \
{\rm color singlet} \ \rangle  = {2\over 3} ~\delta^{ab}.
\label{xixi}
\end{equation}
{}From (\ref{e1}), the amplitude for $i \rightarrow f \pi \pi$ is given by
\begin{equation}
{\cal M} ( i \rightarrow f \pi \pi ) = {2\over 3}~\langle
\ f \ | \ r_{i} \ G_{8} \ r_{j}
\ | \ i \ \rangle ~ \langle \ \pi \pi \ | \ \pi \alpha_{s}~E_{i}^{a}
E_{j}^{a} \ | \ 0 \ \rangle,
\label{amptwo}
\end{equation}
where $G_{8}$ is the Green's function for the color octet $Q\bar{Q}$
states :
\begin{equation}
G_{8} (E) = \sum_{k} ~ { | \ k \ \rangle \ \langle \ k \ | \over
E_{k} - E } .
\label{greenfunction}
\end{equation}
Here $k$ runs over color octet $Q\bar{Q}$ states only.
$E$ and $E_k$ are
the energies of the initial and the intermediate states.
$G_8$ is unknown due to our ignorance of quark confinement in QCD,
and will be treated as a constant.
Then, a lower bound on $G_8$ can be derived
$$
| G_{8} |^{2} > 18~~{\rm GeV}^{-2},
$$
using a sum rule  on $\langle nS \ | \ r^{2} \
| \ 1S \rangle$ and  the decay rate of $\Upsilon (2S) \rightarrow
\Upsilon (1S)
\ \pi \pi$ \cite{ko}.
$G_{8}$ can be determined if the absolute decay rate of $h_{c} (1P)
\rightarrow
J/\psi + \pi^{0}$ is known experimentally \cite{ko}.

Now we show that the $D-$term in (\ref{ampone}) naturally arises from
the $| \Delta L | = 2$ transition, $^{3}D_{1} \rightarrow
^{3}S_{1} \ \pi \pi$.  We assume that the $D-$wave mixes with the initial
quarkonium $\Upsilon (3S)$,
since the final quarkonium $\Upsilon (1S)$ is the lowest level bottomonium
and  it is hard to imagine that it would contain any contamination of
a $D-$wave.
According to potential model calculations \cite{kwongrosner},
there are two $D-$wave levels below $B \bar{B}$ threshold, with
$m(1D) = 10.16~{\rm GeV}$ and $m(2D) = 10.44~{\rm GeV}$, respectively.
Since $\Upsilon (2D)$ is closer to $\Upsilon (3S)$ than $\Upsilon (1D)$,
one might guess the $D-$wave in (\ref{mixing}) to be $\Upsilon (2D)$.
However, as discussed below (\ref{numrr}), a larger value of
$\langle 1S \ | \ r^{2} \ | \ D \rangle $ is desirable for mixing.
Thus, $\Upsilon (1D)$ may be preferred
because it has no node in the radial wave function.   In this work, we
do not address questions regarding the origin of the $S-$ and $D-$wave
mixing and which of the two $D-$wave levels enters in (\ref{mixing}).
The discussion in this section is independent of such issues.
In the next section on radiative transitions of $\Upsilon (3S)$, we consider
both $1D$ and $2D$ mixing.

The angular part of the matrix elements between quarkonia
can be easily performed, and we get  \cite{moxhay}
\begin{eqnarray}
\langle (^{3}S_{1})_{j}^{'} \ | \ r_{k} \ G_{8} \ r_{l} \ | \
(^{3}S_{1})_{i} \rangle & = & {1 \over 3}~I_{S,S^{'}}~
\delta_{ij}~\delta_{kl},
\label{iss}
\\
\langle (^{3}S_{1})_{j} \ | \ r_{k} \ G_{8} \ r_{l} \ | \
(^{3}D_{1})_{i} \rangle & = & {\sqrt{2} \over 10}~I_{S,D}~
\left( \delta_{ik}~\delta_{jl} + \delta_{il}~\delta_{jk}
- {2\over 3}~\delta_{ij}~\delta_{kl} \right),
\label{isd}
\end{eqnarray}
where
\begin{equation}
I_{i,f}  \equiv  \langle f \ | \ r_{k} \ G_{8} \ r_{k} \ | \ i \rangle
= \int_{0}^{\infty}~R_{f} \ r_{k}  \ G_{8} \ r_{k} \ R_{i} ~ r^{2} \ dr.
\label{iff}
\end{equation}

The gluonic matrix element, $\langle \pi \pi \ | \ \pi \alpha_{s}~
E_{i}^{a} E_{j}^{a} \ | \ 0 \rangle$, can be calculated  by considering
$\langle \pi \pi \ | \ \alpha_{s}~G_{\mu \rho}~G_{\nu \sigma} \
| \ 0 \rangle$ and the QCD scale anomaly.
Detailed procedures are discussed in Refs.~\cite{ns}, \cite{v1}
and \cite{moxhay}.
The result can be summarized as
\begin{eqnarray}
&& \langle \pi^{+} \pi^{-} \ | \ \alpha_{s}~
G_{\mu \rho}~G_{\nu \sigma} \
| \ 0 \rangle  \label{ggpipi}   \\
& = &  A~( q^{2} + m_{\pi}^{2} )
+ B \ ( g_{\mu \nu} \tau_{\rho \sigma} - g_{\mu \sigma} \tau_{\rho \nu}
+ g_{\rho \sigma} \tau_{\mu \nu} - g_{\rho \nu} \tau_{\mu \sigma} ),
\nonumber
\end{eqnarray}
where
\begin{equation}
\tau_{\mu \nu}  =  p_{\mu} p_{\nu}^{'} + p_{\nu} p_{\mu}^{'}
= {1\over 2}~( q_{\mu} q_{\nu} - r_{\mu} r_{\nu} ),
\end{equation}
\begin{equation}
A  =  {1 \over 3}~{\alpha_{s}^{2} \over \beta} = - {1 \over 3}~
{2 \pi \over 9},
\label{a}
\end{equation}
\begin{equation}
B  =  {1 \over 2}~\alpha_{s} \rho_{G} = {1 \over 2 \pi}~\lambda,
\label{b}
\end{equation}
and $q = p + p^{'}, ~r = p - p^{'}$.

The $A-$term receives contribution from the QCD scale anomaly \cite{vz},
while the $B-$term arises from the gluonic contribution to the
energy momentum   tensor of QCD \cite{ns}.
The parameter $\lambda$ can be determined from the $\pi \pi$ spectrum in
$\Upsilon (2S) \rightarrow \Upsilon (1S) \ \pi \pi$ \cite{kod}
: $\lambda = 1.6 \sim 1.9$.
This is consistent with what we obtain below  from the
$\pi \pi$ spectrum in $\Upsilon (3S) \rightarrow \Upsilon (1S) \ \pi \pi$.

Using the information given above,
one can calculate the $S-$ and $D-$wave contributions to $\Upsilon (3S)
\rightarrow \Upsilon (1S) \pi \pi$ :
\begin{eqnarray}
{\cal M} ( 3S \rightarrow 1S \ \pi \pi ) & = & {4 \pi^2 \over 81}~I_{3S,1S} ~
\hat{\epsilon} \cdot \hat{\epsilon}^{'} \label{sspipi}
\\
& \times & \left[ q^{2} + m_{\pi}^{2} -
{9 \lambda \over 4 \pi^2}~\left\{
(q^{0})^{2} - (r^{0})^{2} + q^{2} + 2 m_{\pi}^{2} \right\} \right],
\nonumber
\end{eqnarray}
\begin{equation}
{\cal M} ( kD \rightarrow 1S \ \pi \pi )   =
{\sqrt{2}~\lambda \over 30}~I_{kD,1S}
{}~\hat{\epsilon}_{k} \ \hat{\epsilon}_{l}^{'}
{}~\left[ \left\{ q_{k} q_{l} - r_{k} r_{l} \right\} - {1\over 3}~
\delta_{kl} \left\{ \vec{q}^{2} - \vec{r}^{2} \right\} \right],
\label{sdpipi}
\end{equation}
\begin{equation}
{\cal M} ( \Upsilon (3S) \rightarrow \Upsilon (1S) \ \pi \pi )
  =  {\cal M} ( 3S \rightarrow 1S \ \pi \pi )
{}~\cos \phi + {\cal M} ( kD \rightarrow 1S \ \pi \pi ) ~ \sin \phi.
\label{pipitot}
\end{equation}

\noindent
Note that the structure of the $D-$term in (\ref{ampone}) comes from the
first curly bracket in (\ref{sdpipi}), as mentioned at the beginning of
this section.

We fit the $\pi \pi$ spectrum in $\Upsilon (3S) \rightarrow \Upsilon (1S)
\ \pi \pi$ using
the above amplitude with three free parameters, $I_{kD,1S} ~\sin \phi,
{}~~ I_{3S,1S}~\cos \phi$ and $\lambda$.
The best fit is given by two sets of solutions (see Fig.~1) :
\begin{eqnarray}
{I_{kD,1S} \over I_{3S,1S}}~\tan \phi & = & \pm ( 2.4 \pm 0.5 ),
\nonumber   \\
\lambda & = & ( 2.0 \pm 0.1 ),  \label{numone}
\end{eqnarray}
with $\chi^{2} / d.o.f. = 11.2 / 7$ (equivalent to 13.2 \% C.L.).
These correspond to two best fits (called P1 and P2) obtained
in Ref.~\cite{chakrako} using amplitude (\ref{ampone}).
More specifically, one can express (\ref{pipitot}) in the form of
(\ref{ampone}) using (\ref{sspipi}) and (\ref{sdpipi}), and find the
value of $D$.   It turns out that the upper and the lower signs in
(\ref{numone}) correspond to the parameter set P2 and P1 (See table 1.) in Ref.
\cite{chakrako}, respectively.
They can be distinguished by measuring various angular distributions
of the final decay products as suggested  in Ref.~\cite{chakrako}.
Also, as discussed below in detail,
the decay rate for $\Upsilon (3S) \rightarrow \Upsilon (1S) + \eta$
can resolve this twofold ambiguity, and the parameter set P2 is
preferred.
The value of $\lambda = (2.0 \pm 0.1)$ obtained here is consistent with
the $\lambda$ extracted from  the $\pi \pi$ spectrum in $\Upsilon (2S)
\rightarrow \Upsilon (1S) \ \pi \pi$ \cite{kod}.

{}From the
absolute decay rate of $\Upsilon (3S) \rightarrow
\Upsilon (1S) \ \pi \pi$, we obtain
the absolute values of $I_{1D,1S} \sin \phi$ and $I_{3S,1S} \cos \phi$ :
\begin{eqnarray}
| I_{3S,1S}~\cos \phi |  =  | \langle 1S \ | \ r \ G_{8} \ r \
| \ 3S \rangle |~
\cos \phi \ \approx  0.78 ~~{\rm GeV}^{-3}   \label{numiss} \\
| I_{kD,1S}~\sin \phi |  =  | \langle 1S \ | \ r \ G_{8} \ r \
| \ kD \rangle |~
\sin \phi \
\approx  1.92  ~~{\rm GeV}^{-3}  \label{numisd}
\end{eqnarray}
\begin{equation}
\tan \phi  =   2.46 ~{| \langle 1S \ | \ r G_{8} \ r \ | \ 3S \rangle | \over
| \langle 1S \ | \ r \ G_{8} \ r \ | \ kD \rangle |}.
\label{numphi}
\end{equation}
Up to now, we didn't care about $G_8$.  If we assume that $G_{8}$ is a
constant,  (\ref{numphi}) becomes
\begin{equation}
\tan \phi  =   2.46 ~{| \langle 1S \ | \ r^{2} | \ 3S \rangle | \over
| \langle 1S \ | \ r^{2} \ | \ kD \rangle |}.
\end{equation}
Using the values of quarkonia matrix elements
quoted in Ref.~\cite{moxhay},
\begin{eqnarray}
| \langle 1S \ | \ r^{2} \ | \ 3S \rangle | = 0.3 ~~{\rm GeV}^{-2},
\nonumber   \\
| \langle 1S \ | \ r^{2} \ | \ kD \rangle | = 1.65~~{\rm GeV}^{-2},
\label{numrr}
\end{eqnarray}
we get $\phi \approx \pm \ 24^{\circ}$.  However, as discussed in
Ref.~\cite{moxhay}, the matrix element $| \langle 1S \ | \ r^{2} \
| \ 3S \rangle |$ can be much smaller than the above number, since the $3S$
state has  two nodes which may lead to almost complete cancellation.
Furthermore, the accuracy of the wave functions determined in potential models
is about $10 \%$.  Therefore, the actual mixing angle $\phi$ may be {\it much
smaller than} $24^{\circ}$.   Also, because of our approximation on $G_8$,
this kind of determination of $\phi$ is less reliable than that obtained
in the next   section  from radiative decays,
$\Upsilon (3S) \rightarrow \chi_{bJ} (2P) + \gamma$.
In fact,  too large a mixing angle ($\phi$) may result in severe
discrepancy between the theoretical predictions and the experiments
for these  radiative decays.

We can also consider the similar decay $\Upsilon (3S)
\rightarrow \Upsilon (2S) \ \pi \pi$.
The best fit for
$\phi = 0^{\circ}$ (no $D-$wave mixing) yields $\chi^{2} / d.o.f. =
1.7 / 7.$   If the $D-$wave mixing is allowed,  we get the best fit with
$\chi^{2} / d.o.f. = 1.4 / 6$.  Therefore, it is difficult to tell
which of the two is a better fit, and our assumption on the $D-$wave mixing
cannot be tested clearly in this decay mode.  Improved measurements
of the $\pi \pi$ spectrum in $\Upsilon (3S)
\rightarrow \Upsilon (2S) \ \pi \pi$ are welcome.

The message of this work can be put in the following way :
{\it a small mixture of a $D-$wave component in $\Upsilon (3S)$ can explain
the $\pi \pi$ spectrum in $\Upsilon (3S) \rightarrow \Upsilon (1S)
\ \pi \pi$ in the framework of
QCD multipole expansion, if
$| \langle 1S \ | \ r \ G_{8} \ r \ | \ 3S \rangle | $ is much more
suppressed  compared to $| \langle 1S \ | \ r \ G_{8} \ r \ | \ kD
\rangle |$.}

One can also consider another transition, $\Upsilon (3S) \rightarrow
\Upsilon
(1S) + \eta ~~
({\rm or}~~ \pi^{0})$, which occurs through interference of
$E1$ and $M2$ interaction.  The  $M2$ interaction Hamiltonian is
given by \cite{vz}
\begin{equation}
H_{int} (M2) = - {1 \over 4 m}~g \ S_{j}~\xi^{a}~r_{i} D_{i} H_{j}^{a} (0),
\label{m2}
\end{equation}
where  $m$ is the mass of the quark,
$\vec{S}$ is the total spin of the quark and the antiquark,
and $D_{i}$ is the spatial component of the covariant derivative.

{}From (\ref{e1}) and (\ref{m2}), we can derive the amplitude
for this transition and calculate the gluonic matrix element using the
$U_{A}(1)$ anomaly in QCD \cite{vz} :
\begin{eqnarray}
&& {\cal M} ( \Upsilon (3S) \rightarrow \Upsilon (1S) + \eta )   \nonumber  \\
& = & i \left( \partial_{k} \langle \eta \ | \ \pi \alpha_{s}~E_{k}^{a}
H_{j}^{a} \ | \ 0 \rangle \right)~m_{Q}^{-1}~\epsilon_{ijl} \
\hat{\epsilon}_{i}
\ \hat{\epsilon}_{l}^{'}~
\left[ I_{3S,1S}~\cos \phi - {1\over \sqrt{2}} ~I_{kD,1S}~\sin \phi
\right]  \label{sseta}
\\
& =  &{\pi^{2} \over 9}
{}~\left( {3 \over 2} \right)^{1\over 2} f_{\pi} \ m_{\eta}^{2} \ m_{Q}^{-1}
\left[ \epsilon_{ijk} \ \hat{\epsilon}_{i} \ \hat{\epsilon}_{j}^{'} \
(\vec{p})_{k} \right]~
\left[ I_{3S,1S}~\cos \phi  - {1\over \sqrt{2}} ~I_{kD,1S}~\sin \phi
\right],    \nonumber
\end{eqnarray}
where $I_{i,f}$ is defined in (\ref{iff}).
Using the results (\ref{numiss}) and (\ref{numisd}),  we predict
\begin{eqnarray*}
\Gamma ( \Upsilon (3S) \rightarrow \Upsilon (1S) + \eta ) =
\left\{ \begin{array}{cc}
58~~{\rm eV}  & ({\rm for ~~ P2}),    \\
870~~{\rm eV} & ({\rm for ~~ P1}),
\end{array}
\right.
\end{eqnarray*}
or $0.2 \%$ or $3.6 \%$ in the branching ratio.
Current upper limit on this decay mode is $0.22 \%$ \cite{brock},
which prefers the first set (P2) :
\begin{equation}
B ( \Upsilon (3S) \rightarrow \Upsilon (1S) + \eta ) =
0.2 \% ~~~({\rm for ~~P2}),
\label{gsseta}
\end{equation}
which is close to the current upper limit.
Therefore, twofold ambiguity encountered in Ref.~\cite{chakrako} is lifted
in the present work,
and the parameter set P2 is preferred.
Observation of $\Upsilon (3S) \rightarrow \Upsilon (1S)
+ \eta $ at the anticipated branching
ratio would constitute one of the cleanest tests of our assumption :
{\it applicability of QCD multipole expansion to $\Upsilon (3S)$, and
a small admixture of $D-$wave component in $\Upsilon (3S)$.}

\vspace{.4in}
\noindent
{\Large \bf 3. Radiative transitions of $\Upsilon (3S)$ }

\vspace{.2in}

In order to further check the $D-$wave mixing in $\Upsilon (3S)$, we consider
electric dipole radiative transitions, $\Upsilon (3S) \rightarrow
\chi_{bJ} (2P)
+ \gamma$ and
$\Upsilon (3S) \rightarrow \chi_{bJ} (1P)  + \gamma$.
In this section and the following one, we assume the $D-$wave in
$\Upsilon (3S)$ can be either $1D$ or $2D$ state and consider both
possibilities on the same footing.
The transition rate of these decays is given by
\begin{equation}
\Gamma ( \Upsilon (3S) \rightarrow \chi_{bJ} (nP) + \gamma ) =
{4\over 27}~\alpha~Q_{b}^{2}~\omega^{3}~( 2 J_{f} + 1 )~
| \langle nP_{J} \ | \ r \ | \ \Upsilon (3S) \rangle |^{2},
\label{edipole}
\end{equation}
where
\begin{eqnarray}
\langle nP_{0} \ | \ r \ | \ \Upsilon (3S) \rangle
= \langle nP\ | \ r \ | \ 3S
\rangle ~\cos \phi ~ + \sqrt{2} ~\langle nP\ | \ r \ | \ kD \rangle ~\sin \phi,
\nonumber   \\
\vspace{.1in}
\langle nP_{1} \ | \ r \ | \ \Upsilon (3S) \rangle
= \langle nP\ | \ r \ | \ 3S
\rangle ~\cos \phi ~ - {1\over \sqrt{2}} ~\langle nP\ | \ r \ |
\ kD \rangle ~\sin \phi,   \label{spr}
\\
\langle nP_{0} \ | \ r \ | \ \Upsilon (3S) \rangle
= \langle nP\ | \ r \ | \ 3S
\rangle ~\cos \phi ~ + {1 \over 5 \sqrt{2}} ~\langle nP\ | \ r \
| \ kD \rangle ~\sin \phi .
\nonumber
\end{eqnarray}
The measured rates of $\Upsilon (3S) \rightarrow \chi_{bJ} (2P)
+ \gamma$  are available
\cite{pdg} for $J=0,1,2$.
Analysis of these decay rates yield a solution for
the two variables $\langle 2P|r|3S\rangle ~\cos\phi$ and
$\langle 2P|r|kD\rangle ~\sin\phi$ :
\begin{eqnarray}
\langle 2P\ | \ r \ | \ 3S \rangle ~\cos \phi & = &
                      +(2.66 \pm 0.16)~~{\rm GeV}^{-1}, \nonumber  \\
\langle 2P\ | \ r \ | \ kD \rangle ~\sin \phi & = &
                            -(0.14 \pm 0.18)~~{\rm GeV}^{-1}.  \label{dspr}
\end{eqnarray}

To determine $\phi$ from (\ref{dspr}), we use the potential model calculations
of $\langle \ f \ | \ r \ | \ i \ \rangle$ given in \cite{kwongrosner} :
\begin{eqnarray}
\langle 1P \ | \ r \ | \ 3S \rangle & =  -0.023~~&{\rm GeV}^{-1},
\nonumber \\
\langle 1P \ | \ r \ | \ 1D \rangle & =  -2.0~~  &{\rm GeV}^{-1},
\nonumber \\
\langle 1P \ | \ r \ | \ 2D \rangle & =  -0.26~~ &{\rm GeV}^{-1}
\nonumber \\
\langle 2P \ | \ r \ | \ 3S \rangle & =  +2.7~~  &{\rm GeV}^{-1},
\label{numspr}  \\
\langle 2P \ | \ r \ | \ 1D \rangle & =  +1.9~~  &{\rm GeV}^{-1},
\nonumber \\
\langle 2P \ | \ r \ | \ 2D \rangle & =  -2.7~~  &{\rm GeV}^{-1}.
\nonumber
\end{eqnarray}

Multiplicative relativistic correction factors to the matrix elements
involving the $S-$wave have been calculated in Ref. \cite{moxrosner}. The
corrections depend on the state of $\chi_{bJ}(nP)$ into which
the $\Upsilon (3S)$ decays :
\begin{eqnarray}
\langle 2P_{J} \ | \ r \ | \ 3S \rangle : \qquad
                                         & J=2 &\qquad 1.02,  \nonumber \\
                                       ~ & J=1 &\qquad 1.00,  \nonumber \\
                                       ~ & J=0 &\qquad 1.95.  \label{rela} \\
\langle 1P_{J} \ | \ r \ | \ 3S \rangle : \qquad
                                         & J=2 &\qquad 2.3,  \nonumber \\
                                       ~ & J=1 &\qquad 1.2,  \nonumber \\
                                       ~ & J=0 &\qquad 1.9.  \nonumber
\end{eqnarray}

{}From the second equation of (\ref{dspr}) and (\ref{numspr}),  we get a value
for the $D-$wave mixing angle $\phi$ as deduced from $\Upsilon (3S)$ decaying
into $\chi_{bJ}(2P)$ :
\begin{eqnarray}
\phi_{1D} = -4^\circ \pm 6^\circ \nonumber \\
\phi_{2D} = +3^\circ \pm 4^\circ \label{ang}
\end{eqnarray}
The error in the angles is an estimate only. Moreover,
the angles are seen to be consistent with zero.
$\phi$ may also be estimated from the first equation of (\ref{dspr}) and
Eqs. (\ref{numspr}) and (\ref{rela}). $-12^\circ <\phi <+12^\circ$ obtained
this way is not very useful because the allowed range is large.

A determination of the mixing angle is also possible by making use of
experimental bounds on various combinations of branching ratios for
$\Upsilon (3S)$ decaying into $\chi_{bJ}(1P)$. Form Ref. \cite{pdg} we have
the experimentally measured
\begin{eqnarray}
B(\chi_{b2}(1P)\rightarrow \Upsilon (1S)\gamma )& = & 0.22\pm 0.04 \nonumber \\
B(\chi_{b1}(1P)\rightarrow \Upsilon (1S)\gamma )& = & 0.35\pm 0.08
                                                               \label{hein1} \\
B(\chi_{b0}(1P)\rightarrow \Upsilon (1S)\gamma )& < & 0.06.   \nonumber
\end{eqnarray}

\noindent
$B(\Upsilon (3S)\rightarrow\chi_{bJ}(1P)\gamma)$ is sensitive to the $D-$wave
mixing angle and can be calculated from Eq. (\ref{edipole}), (\ref{numspr})
and (\ref{rela}) knowing the total decay width.
These branching ratios lead to a bound on $\phi$ because one has
to satisfy the following experimental relation \cite{heintz} (individual
branching ratios are not available at this time) :
\begin{eqnarray}
F(\phi ) =\sum_{J=1,2}
               B(\Upsilon (3S)&\rightarrow&\chi_{bJ}(1P)\gamma )
         \cdot B(\chi_{bJ}(1P)\rightarrow\Upsilon (1S)\gamma ) \nonumber \\
          &= & (1.2 \ ^{+0.4}_{-0.3} \pm 0.09)\times 10^{-3}.   \label{hein2}
\end{eqnarray}
In Figs.~2 (a) and (b) we show $F(\phi )$ for  mixing with $|1D\rangle$
and $|2D\rangle$ states respectively. The allowed region of $\phi$ is larger
in the case of $2D$ mixing if one demands the consistency between the
$\chi_{bJ} (nP)$ decays.
We like to emphasize that  the present experimental data is
consistent with the assumption of a $D-$wave mixing. Based on our analysis,
mixing with the $2D$ state is seen to be more plausible.

An estimate of the mixing angle (\ref{ang}) may now be used to predict the
branching ratio of $\chi_{b0}$ decaying into $\Upsilon (1S)$ using
\cite{heintz}
\begin{eqnarray}
\sum_{J=0,1,2} B(\Upsilon (3S) &\rightarrow &\chi_{bJ}(1P)\gamma )
         \cdot B(\chi_{bJ}(1P)\rightarrow\Upsilon (1S)\gamma ) \nonumber \\
             & = &(1.7 \pm 0.4\pm 0.6)\times 10^{-3}~.         \label{hein3}
\end{eqnarray}

We finally write down expected branching ratios of $\Upsilon (3S)
\rightarrow \chi_{bJ} (1P) + \gamma$ for $J = 0, 1, 2$
assuming $|1D\rangle$ and  $|2D\rangle$ mixing using the central values
of Eq. (\ref{ang}):
\begin{eqnarray}
B ( \Upsilon (3S) \rightarrow \chi_{b0} (1P) + \gamma )
               & = 1.1~\% ~~(1D) ,\qquad &1.8~\% ~~(2D),  \nonumber \\
B (  \Upsilon (3S) \rightarrow \chi_{b1} (1P) + \gamma )
               & = 2.0~\% ~~(1D) ,\qquad &0.2~\% ~~(2D),  \label{exbr} \\
B ( \Upsilon (3S) \rightarrow \chi_{b2} (1P) + \gamma )
               & = 0.3~\% ~~(1D) ,\qquad &1.0~\% ~~(2D),  \nonumber
\end{eqnarray}
It is clear that a better determination of branching ratios of these radiative
decays, $\Upsilon (3S) \rightarrow \chi_{bJ} (nP) + \gamma$ with $n = 1, 2$
can resolve $2D$ mixing from $1D$ mixing, or vice versa.

The spin--flip radiative transition $\Upsilon (3S) \rightarrow
\eta_{b} + \gamma$
is also affected by the $D-$wave component in $\Upsilon (3S)$.
The decay rate is given by
\begin{equation}
\Gamma ( \Upsilon (3S) \rightarrow \eta_{b} + \gamma )
= {2 \over 3}~\alpha~Q_{b}^{2}~
{\omega^3 \over m_{b}^2}~| F |^{2},
\label{sebg}
\end{equation}
where
\begin{equation}
F  = \sqrt{2}~\langle 1S | \ j_{0} ( \omega r / 2 ) \ | 3S
\rangle ~ \cos \phi
+ ~\langle 1S | \ j_{2} ( \omega r / 2 ) \ | kD
\rangle ~ \sin \phi,
\label{ff}
\end{equation}
where $j_{n} (x)$ is the $n-$th spherical Bessel function.
In the long wavelength limit ($\omega r \rightarrow 0$),
we can approximate:
\begin{eqnarray*}
j_{0} (x) & = & 1 - {x^{2} \over 6},   \\
j_{2} (x) & = & {x^{2} \over 15},
\end{eqnarray*}
so that
\begin{equation}
F = {1\over G_8}~\left[ - {\sqrt{2} \over 24}~I_{3S,1S}~\cos \phi
+ {1 \over 60}~I_{kD,1S}~\sin \phi \right],
\end{equation}
assuming $G_8$ is a constant parameter, as in Ref.~\cite{ko}.
Therefore, one can again use (\ref{numiss}) and (\ref{numisd}) to evaluate
$F$, and calculate the decay rate from  (\ref{sebg}) :
\begin{eqnarray}
\Gamma ( \Upsilon (3S) \rightarrow \eta_{b} + \gamma ) =
\left\{ \begin{array}{c}
4~\left( 18 / G_{8}^{2} ({\rm GeV}^{-2}) \right)
                        ~~{\rm eV}~~~{\rm (for ~~P2)},   \\
75~\left( 18 / G_{8}^{2} ({\rm GeV}^{-2}) \right)
                        ~~{\rm eV}~~~{\rm (for ~~P1)},
\end{array}
\right.
\label{numsebg}
\end{eqnarray}
Note that (i) there is no $\phi-$dependence left over, once we use the results
(\ref{numiss}) and (\ref{numisd}), and (ii) this result is independent of
which of the $D$-wave actually mixes.
If the parameter set P1 were the correct one and $G_8$ not too large,
this could open up a new option for the discovery of  $\eta_b$ in the
$e^+ e^-$ annihilation  bypassing the intermediate
stage involving $h_{b} (1P)$.
Unfortunately, the upper limit on $B ( \Upsilon (3S) \rightarrow
\Upsilon (1S) +
\eta )$ prefers the parameter set P2, for which  $B (
\Upsilon (3S) \rightarrow \eta_{b}
+ \gamma ) < 1.6 \times 10^{-4}$.
Therefore, this channel may compete with other possibilities discussed below
in  the search of $\eta_b$, only if $G_{8}^{2}$ is not too large.

The $D-$wave component in $\Upsilon (3S)$ can generate other interesting
radiative transitions :
$$
\Upsilon (3S) \stackrel{\gamma}{\longrightarrow}~1^{1}D_{2}
\stackrel{\gamma}{\longrightarrow}~
h_{b} (1P) \stackrel{\gamma}{\longrightarrow}~\eta_{b}.
$$
The first chain is energetically allowed only for $1D$ mixing.
These decay rates can be readily obtained from the results of
Ref.~\cite{novikov}.  Omitting all details, the final results are given
below :
\begin{eqnarray}
\Gamma ( \Upsilon (3S) \rightarrow
1^{1}D_{2} + \gamma ) & = & {16 \over 3}~\omega^{3}~
{\alpha ~Q_{b}^{2} \over (2 m_{b})^2}~\sin^{2} \phi
\label{sdg}
\\
& \approx &
3.8 \left( {\omega ({\rm MeV}) \over 200} \right)^{3}~\left( { \sin \phi
\over 0.1} \right)^{2}~~~{\rm eV},
\nonumber
\end{eqnarray}
\begin{equation}
\Gamma ( 1^{1}D_{2} \rightarrow h_{b} (1P) + \gamma ) = {8
\over 15}~\alpha ~Q_{b}^{2}~
\omega^{3}~| \langle 1P \ | \ r \ | \ 1D \rangle |^{2}
\approx  30~~{\rm keV},
\label{dhg}
\end{equation}
\begin{equation}
\Gamma ( h_{b} (1P) \rightarrow \eta_{b} + \gamma )  \approx  40~~{\rm keV}.
\label{hebg}
\end{equation}
This corresponds to the production of $\sim 80~h_{b} (1P)$'s in  decays of
$10^{6}~\Upsilon (3S)$'s for $\sin \phi = \pm 0.1$, or $\phi = \pm 6^{\circ}$,
and if $B ( 1^{1}D_{2} \rightarrow h_{b} (1P) + \gamma ) = 50 \%$.
These are sensitive to the mixing angle $\phi$, and may be useless for
producing $h_{b} (1P)$ and $\eta_b$, if $\sin \phi < 0.1$.

\vspace{.4in}
\noindent
{\Large \bf 4. Hadronic transitions of $\Upsilon (3S)$ into $h_{b}
(1P)$ }

\vspace{.2in}

Finally, let us consider $\Upsilon (3S) \rightarrow h_{b} (1P) + X$
with $X = \pi^{0}$ or
$\pi \pi$. This may serve
as a source of the spin--singlet $P-$wave state,
$h_{b} (1P)$, if its branching ratio
is appreciable.
In QCD multipole expansion, the above transitions are generated by
the interference between $E1$ and $M1$ interactions, where the $M1$ interaction
Hamiltonian is \cite{v1}
\begin{equation}
H_{int} (M1) = {1 \over 2 m}~g \ \xi^{a}~\Delta_{i} H_{i}^{a} (0),
\label{m1}
\end{equation}
with $\vec{\Delta}$ being the difference of the spin operators for the
quark and the antiquark.
The amplitude for $\Upsilon (3S) \rightarrow h_{b} (1P) + X$ is
\begin{eqnarray}
{\cal M} (\Upsilon (3S) \rightarrow h_{b} (1P) + X) =
- ~{2\over 3 m}~\hat{\epsilon}_{i}~
\hat{\epsilon}_{j}^{'}~\langle \ X \ | \
\pi \ \alpha_{s} E_{l}^{a}~H_{k}^{a} \ | \ 0 \rangle \nonumber  \\
\times \langle \ (h_{b} (1P))_{j} \ | \ r_{l}\ G_{8} \ \Delta_{k} \ + \
\Delta_{k} \ G_{8} \ r_{l} \ | \ (\Upsilon (3S))_{i} \ \rangle,
\label{amphbx}
\end{eqnarray}
where $\hat{\epsilon}$ and $\hat{\epsilon}^{'}$ are the polarization vectors of
$\Upsilon (3S)$ and $h_{b} (1P)$, respectively.

The angular part of the quarkonium matrix elements can be performed as
before, and we get
\begin{eqnarray}
\langle \ (^{1}P_{1})_{j} \ | \ r_{l}\ G_{8} \ \Delta_{k} \ | \
(^{3}S_{1})_{i} \ \rangle  & = & {1 \over \sqrt{3}}~\langle R_{P} | \ r \ G_{8}
\ | \ R_{S} \rangle ~\delta_{ik}~\delta_{jl},
\label{prs}
\\
\langle \ (^{1}P_{1})_{j} \ | \ r_{l}\ G_{8} \ \Delta_{k} \ | \
(^{3}D_{1})_{i} \ \rangle  & = & {\sqrt{3} \over 5 \sqrt{2}}~
\langle R_{P} | \ r \ G_{8} \ | \ R_{D} \rangle
\nonumber    \\
& & \times ~\left(
\delta_{ij}~\delta_{kl} \ + \ \delta_{il}~\delta_{jk} \ - \
{2\over 3}~\delta_{ik}~\delta_{jl} \right).
\label{prd}
\end{eqnarray}

Now, consider the case $X = \pi^0$, for which the matrix element
of the gluonic operators are determined by $U_{A}(1)$ anomaly and the
mass difference between $u$ and $d$ quarks \cite{v1} \cite{vz} :
\begin{equation}
\langle \pi^{0} \ | \ \pi ~\alpha_{s}~ E_{i}^{a}~H_{k}^{a} \ | \
0 \rangle = \delta_{ik}~{\pi^{2} \over 3 \sqrt{2}}~\left( { m_{u} -
m_{d} \over m_{u} + m_{d}} \right)~ f_{\pi}~m_{\pi}^{2}.
\equiv A_{0}~\delta_{ik}.
\end{equation}
Using the pion decay constant $f_{\pi} = 132$ MeV and
$( m_{u} - m_{d} ) / ( m_{u} +  m_{d} ) =
0.3$, we get $A_{0} = 1.7 \times 10^{-3}~~{\rm GeV}^3$.
The amplitude for $\Upsilon (3S) \rightarrow h_{b} (1P) + \pi^0$ becomes
\begin{equation}
{\cal M} (\Upsilon (3S) \rightarrow h_{b} (1P) + \pi^0) = A_{0}
{}~I_{\pi}~\hat{\epsilon}
\cdot \hat{\epsilon}^{'},
\end{equation}
where
\begin{equation}
I_{\pi} = - {4 \sqrt{3} \over 9}~{G_{8} \over m_{b}} ~\left[ ~\langle 1P
\ | \ r \
| \ 3S \rangle~\cos \phi~ +  ~ \sqrt{2}~\langle 1P \ | \ r \
| \ 1D  \rangle ~\sin \phi \ \right],
\label{ipi}
\end{equation}
and the decay rate is
\begin{equation}
\Gamma ( \Upsilon (3S) \rightarrow h_{b} (1P) + \pi^{0} )  =  {1 \over 2 \pi}~
( A_{0}\ I_{\pi} )^{2}
{}~| \vec{p}_{\pi} |.
\label{upstohb}
\end{equation}
The expression in the bracket of (\ref{ipi}) is the same
as the first one in (\ref{spr}).
Therefore, $\Upsilon (3S) \rightarrow \chi_{b0} (1P) + \gamma$ and
$\Upsilon (3S) \rightarrow h_{b} (1P) + \pi^0$ are related with each other.
The ratio of the decay rates of these two seemingly different
decays is independent of quarkonium matrix
elements of $r$ or the mixing angle $\phi$, and determines $G_8$.
{}From (\ref{edipole}) and (\ref{upstohb}),  we find
\begin{eqnarray}
{\Gamma (\Upsilon (3S) \rightarrow h_{b} (1P) + \pi^{0} ) \over
\Gamma (\Upsilon (3S) \rightarrow \chi_{b0} (1P) + \gamma)} & = &
{27 \over 8 \pi \alpha Q_{b}^{2}}~\left( {4 \sqrt{3} \over 9}~{G_{8} A_{0}
\over m_{b} \omega } \right)^{2}
\nonumber   \\
& \approx & 8.7 \times 10^{-3}~
\left( { G_{8}^{2}~
({\rm GeV}^{-2}) \over 18 } \right).
\label{ratio}
\end{eqnarray}
The absolute decay rate of $\Upsilon (3S) \rightarrow h_{b} (1P) + \pi^{0}$ is
\begin{eqnarray}
\Gamma ( \Upsilon (3S) \rightarrow h_{b} (1P) + \pi^{0} )  =
\left( {G_{8}^{2}~({\rm GeV}^{-2}) \over 18 } \right) \times
\left\{  \begin{array}{cc}
1~~{\rm eV} & ({\rm for} ~~1D)   \\  0.3~~{\rm eV} & ({\rm for} ~~2D)
\end{array}   \right.
\label{gupstohb}
\end{eqnarray}
(For numerical estimates, we have used the values of $\langle 1P | r
| 3S \rangle$ and $\langle 1P | r | 1D \rangle$ quoted in (\ref{numspr}) with
relativistic correction factors (\ref{rela}) and central values of
(\ref{ang}).)
This amounts to the branching ratio greater than $4 \times 10^{-5}$ for
$1D$ mixing, and $ 1.2 \times 10^{-5}$ for $2D$ mixing.
Therefore,  this decay may be the best for reaching
$h_{b} (1P)$, and subsequently $\eta_b$ through $h_{b} (1P)
\rightarrow \eta_{b} + \gamma$.

A similar decay, $\Upsilon (3S) \rightarrow h_{b} (1P) \pi \pi$,
does not receive any contribution
from the trace of the energy--momentum tensor in QCD, and is not enhanced
over $\Upsilon (3S) \rightarrow h_{b} (1P) + \pi^{0}$.
{}From the general expression (\ref{ggpipi})--(\ref{b}), we get
\begin{equation}
\langle \ \pi^{+} \pi^{-} \ | \ \pi \ \alpha_{s} \ E_{i}^{a}~H_{j}^{a}
\ | \ 0 \rangle = {1 \over 2}~\lambda~\epsilon_{ijk}~( E_{1} p_{2 k} +
E_{2} p_{1 k} ),
\end{equation}
where $p^{\mu} = ( E_{1}, \vec{p}_{1} ), p^{' \mu} = ( E_{2}, \vec{p}_{2} )$
are the four--momenta of the pions.
Then, the decay rate for $\Upsilon (3S) \rightarrow h_{b} (1P)
\pi^{+} \pi^-$ is \cite{v1}
\begin{equation}
\Gamma ( \Upsilon (3S) \rightarrow h_{b} (1P) \pi^{+} \pi^- ) =
{\lambda^{2} \over 48 \pi^3}~\varphi~{\Delta^{7} \over 70}~| I_{2 \pi}
|^{2},
\label{upstohbpipi}
\end{equation}
where
$\Delta = m(\Upsilon (3S)) - m(h_{b} (1P))$,  $\varphi = 0.22$ is the
suppression factor of the phase space integral due to the pion
mass \cite{v1}, and
\begin{equation}
I_{2 \pi} =  - {4 \sqrt{3} \over 9}~{G_{8} \over m_{b}} ~\left[ ~\langle
1P
\ | \ r \
| \ 3S \rangle~\cos \phi~ - ~ {1 \over \sqrt{2}}~\langle 1P \ | \ r \
| \ 1D  \rangle ~\sin \phi \ \right].
\label{i2pi}
\end{equation}
Again, the expression in the square bracket is the same as the second
equation of (\ref{spr}), and this decay is related to
$\Upsilon (3S) \rightarrow \chi_{b1} (1P) + \gamma$
in the same way as in (\ref{ratio}).

{}From (\ref{upstohb}) and (\ref{upstohbpipi}), we get
\begin{eqnarray}
{\Gamma (\Upsilon (3S) \rightarrow h_{b} (1P) \ \pi \pi) \over
\Gamma (\Upsilon (3S) \rightarrow h_{b} (1P) + \pi^{0} )}
\approx
\left\{  \begin{array}{c}
0.2  ~~~({\rm for} ~~1D)\\ 0.03 ~~~({\rm for}~~2D) \end{array}    \right.
\label{gratiopipipi}
\end{eqnarray}
where we have used $\lambda = 2$ and Eqs.
(\ref{numspr}),  (\ref{rela}) and (\ref{ang}).
This corresponds to $\sim 10^{-5}$ (for the $1D$ mixing) and $\sim 4 \times
10^{-7}$ (for the $2D$ mixing) in the branching ratio of $\Upsilon
(3S) \rightarrow
h_{b} (1P) \ \pi \pi$.
The current upper limit is about two orders of magnitude above our prediction
\cite{brock}.
Our analysis shows that one has to look for $\Upsilon (3S) \rightarrow
h_{b}
(1P) + \pi^{0}$
($10^{-5} \sim 10^{-4}$ in the branching ratio)
rather than $\Upsilon (3S) \rightarrow h_{b} (1P) \pi \pi$ to
identify the spin--singlet $P-$wave
bottomonium state, $h_{b} (1P)$, even if $\Upsilon (3S)$
has an admixture of a $D-$wave.

\vspace{.4in}
\noindent
{\Large \bf 5. Conclusion}

\vspace{.2in}

In conclusion, we find that the $\pi \pi$ spectrum in $\Upsilon (3S)
\rightarrow
\Upsilon (1S) \ \pi \pi$ can be explained in a natural way in the framework of
QCD multipole expansion by assuming the physical $\Upsilon (3S)$ state is
an admixture of the $S-$ and $D-$waves.
The $\pi \pi$ spectrum determines the mixing angle $\phi$ to be less
than $\sim 24^{\circ}$. (Decay rates for $\Upsilon (3S) \rightarrow
\chi_{bJ} (2P)
+ \gamma$ give much tighter and more reliable value for $\phi$
to be $- ( 4^{\circ} \pm  6^{\circ} )$ for $1D$ mixing,
and $+( 3^{\circ} \pm 4^{\circ} )$ for  $2D$ mixing.)
Effects of this $D-$wave component on other decays of $\Upsilon (3S)$ are
discussed in detail.
Twofold ambiguity encountered in Ref. \cite{chakrako} is resolved by
the upper limit on $B ( \Upsilon (3S) \rightarrow \Upsilon (1S) +
\eta )$, and the parameter
set P2  is preferred  for which
\[
B ( \Upsilon (3S) \rightarrow \Upsilon (1S) + \eta ) = 0.2 \%.
\]
This is one of the cleanest tests of our assumptions on the $D-$wave mixing
in $\Upsilon (3S)$ and applicability  of QCD multipole expansion
to hadronic transitions of $\Upsilon (3S)$.
Consistency of our approach can be cross--checked by measuring
the polar angle distirbution ( with respect to  the
$e^+ e^-$ beam direction )  of $\Upsilon (1S)$
in $\Upsilon (3S) \rightarrow \Upsilon (1S)
\ \pi \pi$   as discussed in Ref. \cite{chakrako}.
Further tests of our assumptions will be possible by
checking our predictions (\ref{exbr}) by improving measurements of
$B (\Upsilon (3S) \rightarrow \chi_{bJ} (nP) + \gamma)$.
More informations on $\phi$ and/or $G_8$ can be extracted  by checking
(\ref{numsebg}), (\ref{sdg}),
(\ref{dhg}), (\ref{ratio})  and (\ref{gupstohb}).
The most promising place to look for $h_{b} (1P)$ may be $\Upsilon (3S)
\rightarrow h_{b} (1P) + \pi^0$,  but the branching ratio is rather small,
(\ref{gupstohb}).  (If $\sin \phi \approx 0.1$ and $1D$ mixing is the correct
one, the cascades (3) may be  useful.)  Because of the small branching ratio
of decay modes involving $h_{b} (1P)$ in the final state, it may be worth
while to look at the direct
transition $\Upsilon (3S) \rightarrow \eta_{b} + \gamma$ for $\eta_{b}$
search.

In this work, we have derived several new results on the branching ratios of
hadronic and radiative transitions of $\Upsilon (3S)$.
If any of our predictions is in serious contradiction with the experiments,
then we may conclude that QCD multipole expansion breaks down for
hadronic transitions of $\Upsilon (3S)$  to lower quarkonia because of
the large radius of $\Upsilon (3S)$ and its being close to the $B\bar{B}$
threshold.

\vspace{.4in}
\noindent
{\Large \bf Acknowledgements}

\vspace{.2in}

We are grateful to Y. Kubota, R. Poling, S. Rudaz  and M.B. Voloshin for
useful discussions and comments.     P.K. thanks J.L. Rosner for
comments on the quarkonium matrix elements.

This work is supported in  part by the DOE grant DE--AC02--83ER--40105.
S.C. would like to thank the Mechanical Engineering department
for financial support.

\vspace{.8in}

\newpage

\noindent
{\Large \bf Figure Caption}

\vspace{.2in}

{\bf Fig.~1}  The best fit to the $\pi \pi$ spectrum in $\Upsilon (3S)
\rightarrow \Upsilon (1S)
\ \pi \pi$ using the amplitude (\ref{pipitot}).  The results of this fit
is given in (\ref{numone}), for which $\chi^{2} / d.o.f. = 11.2 / 7$.

\vspace{.2in}

{\bf Fig.~2}  Plots of $F(\phi )$ of Eq. (\ref{hein2}) assuming the
mixing angle ($\phi$) to be a parameter. Fig.~2 (a) assumes a mixing with
$|1D\rangle$, while Fig.~2 (b)  assumes a mixing with $|2D\rangle$.
The shaded region is allowed by Eqs. (\ref{ang}) and (\ref{hein2}).
These may be used to obtain bounds on $\phi$.

\vspace{.2in}
\centerline{\bf Table~1.}
\vspace{.1in}
\noindent
\centerline{Two sets of parameters giving the best $\chi^2$ fit
            to the $\pi \pi$ spectrum from Ref. \cite{chakrako}.}
\vspace{.1in}
$$
\vbox{\tabskip 1em plus 2em minus .5em
\halign to 250pt{\hfil #\hfil && \hfil #\hfil \cr
\noalign{\hrule}\noalign{\hrule}
{}~ &~ &~\cr
Parameters    & Fit 1 (P1) & Fit 2 (P2) \cr
{}~ &~ &~\cr
                                               \noalign{\hrule}
{}~ &~ &~\cr
A & $ 101.60 \pm 103.7 $ & $ 366.34 \pm 61.94$ \cr
B & $ -5.80 \pm 8.29 $ & $ -3.12 \pm 2.67$ \cr
C & $ 20.53 \pm 84.09$ & $ 4.33 \pm 24.86 $ \cr
D & $ 3.73 \pm 4.35 $ & $ -1.03 \pm 0.30 $ \cr
{}~ &~ &~\cr
                                \noalign{\hrule}\noalign{\hrule}
}}$$

\end{document}